 \documentclass[preprint,onecolumn,amsmath,amssymb,aps,prb]{revtex4-1}

\usepackage{graphicx}
\usepackage{dcolumn}
\usepackage{bm}
 \usepackage{float} 

\usepackage{fullpage}
\usepackage{color}
\usepackage{subcaption}
\usepackage{multirow}
\usepackage{amsmath}
\usepackage{amssymb}
\usepackage{amsthm}
\usepackage[colorlinks   = true, urlcolor     = cyan, linkcolor    = blue, citecolor   = red]{hyperref}
\usepackage{bm}
\usepackage{pict2e}
\usepackage{epstopdf}
\usepackage{comment}
\usepackage{epsfig}
\usepackage{url}
\usepackage[ruled,vlined]{algorithm2e}

\usepackage{cancel}
\usepackage{amsmath}
\usepackage{amssymb}
\usepackage{etoolbox}
\usepackage{graphicx}
 \usepackage[english]{babel}
\usepackage{mathrsfs} 
\usepackage{enumitem}                                  

\usepackage{amsthm} 
\usepackage{siunitx}
\usepackage{ upgreek }
 \newcommand{\noop}[1]{}

\begin{document}


\title{Strain engineering of Janus transition metal dichalcogenide nanotubes: An ab intio study}

\author{Arpit Bhardwaj}
\affiliation{College of Engineering, Georgia Institute of Technology, Atlanta, GA 30332, USA}

\author{Phanish Suryanarayana}
\email{phanish.suryanarayana@ce.gatech.edu}
\affiliation{College of Engineering, Georgia Institute of Technology, Atlanta, GA 30332, USA}


\begin{abstract}
We study the electromechanical response of Janus transition metal dichalcogenide (TMD) nanotubes from first principles. In particular, considering both armchair and zigzag variants of twenty-seven select Janus TMD nanotubes, we determine the change in bandgap and charge carriers' effective mass upon axial and torsional deformations using density functional theory (DFT). We observe that metallic nanotubes remain unaffected, whereas the bandgap in semiconducting nanotubes decreases linearly and quadratically with axial and shear strains, respectively, leading to semiconductor--metal transitions. In addition, we find that there is a continuous decrease and increase in the effective mass of holes and electrons with strains, respectively, leading to n-type--p-type semiconductor transitions. We show that this behavior is a consequence of  the rehybridization of orbitals, rather than charge transfer between the atoms. Overall, mechanical deformations form a powerful tool for tailoring the electronic response of semiconducting Janus TMD nanotubes.
\end{abstract}

\keywords{Janus transition metal dichalcogenide nanotubes, Density functional theory, Strain engineering, Axial deformation, Torsional deformation}

\maketitle
\section{Introduction}
The synthesis of carbon nanotubes over three decades ago \cite{iijima1991helical} marked an important milestone for nanoscience and nanotechnology, with more than thirty different nanotubes synthesized to date \cite{tenne2003advances, rao2003inorganic, serra2019overview}. These quasi 1D hollow nanostructures demonstrate enhanced and novel electronic, thermal, mechanical, and optical properties compared to their bulk analogues, making their study an active area of research. Given the relatively high correlation between the successful synthesis of nanotubes and their 2D counterparts --- first principles calculations have predicted thousands of atomic monolayers to be stable  \cite{haastrup2018computational, zhou20192dmatpedia, gjerding2021recent} --- a number of new nanotubes with  fascinating properties are likely to be synthesized in the not too distant future.

The transition metal dichalcogenide (TMD) group of nanotubes --- materials denoted by MX\textsubscript{2}, where M and X are used to represent a transition metal and chalcogen, respectively --- is the most varied set, with the highest number of distinct nanotubes synthesized thus far \cite{tenne2003advances, rao2003inorganic, serra2019overview}. However, these nanotubes are generally multi-walled with large diameters --- rationalized by the need for relatively high energies to bend their 2D material analogues \cite{kumar2020bending} --- limiting the appearance of unique and fascinating properties that are typically associated with quantum confinement effects. Furthermore, only a small percentage of all the potential TMD nanotubes have been synthesized thus far, in significant part to the nanotubes generally being  energetically less favorable relative to their 2D counterparts.

Janus TMD nanotubes \cite{yagmurcukardes2020quantum} --- materials denoted by MXY,  where X and Y are used to represent two different chalcogens ---  do not suffer from the aforementioned limitations. In particular, the asymmetry in the system makes the rolled nanotube configuration energetically more  favorable than the corresponding flat sheet \cite{xiong2018spontaneous, bhardwaj2021elastic} --- MoSSe and WSSe monolayers have recently been synthesized \cite{lu2017janus, zhang2017janus, trivedi2020room, lin2020low} ---  significantly increasing the likelihood of  single-walled small-diameter nanotubes with exotic properties/behavior . Therefore, at the very least, it is to be expected that Janus TMD nanotubes have similarly many applications as their non-Janus counterparts, including photodetectors \cite{tang2018janus, oshima2020geometrical, xie2021theoretical, ju2021tuning, ju2021rolling, zhang2019mosse}, nanoelectromechanical (NEMS) devices \cite{yudilevichself, levi2015nanotube, divon2017torsional}, biosensors \cite{barua2017nanostructured}, mechanical sensors \cite{li2016low, sorkin2014nanoscale, oshima2020geometrical}, and superconductive materials \cite{nath2003superconducting, tsuneta2003formation}. 

There have been a number of ab initio studies to characterize the properties of Janus TMD nanotubes \cite{wu2018tuning, mikkelsen2021band, zhao2015ultra, tao2018band, xie2021theoretical, evarestov2020first, bolle2021structural} and their electronic response to mechanical deformations \cite{oshima2020geometrical, wang2018mechanical, tang2018janus, luo2019electronic}. However, these investigations have been restricted to relatively few nanotubes, particularly in the case of electromechanical response, where only MoSSe has been studied to date. Even then, only the electronic response to axial deformations has been studied, with torsional deformations not considered. Also, apart from Ref.~\cite{oshima2020geometrical}, the equilibrium diameter for the nanotube has not been considered in determining the electromechanical response. Overall, the electronic response of Janus TMD nanotubes to axial and torsional deformations has not been studied heretofore, which provides the motivation for the current effort.

In this work, we perform a comprehensive investigation of the electromechanical response of Janus TMD nanotubes to axial and torsional deformations using ab initio simulations.  In particular, considering both armchair and zigzag variants of twenty-seven select Janus TMD nanotubes, we perform Kohn-Sham DFT \cite{hohenberg1964inhomogeneous, kohn1965self} calculations to determine the change in bandgap and charge carriers' effective mass upon deformation, starting from their equilibrium diameters. We find that while metallic nanotubes remain unaffected, the electronic response of semiconducting Janus TMD nanotubes can be tuned by axial and torsional deformations, leading to semiconductor--metal and n-type--p-type semiconductor transitions.


\section{Systems and methods} \label{Sec:Methods}

We consider the following single-walled armchair and zigzag Janus TMD nanotubes at their equilibrium diameters \cite{bhardwaj2021elastic}: (i) M$=$\{Ti, Zr, Hf\}, X $=$\{S, Se, Te\}, and Y$=$\{S, Se, Te\}, with 1T-o symmetry; and (ii) M$=$\{V, Nb, Ta, Cr, Mo, W\}, X$=$\{S, Se, Te\}, and Y$=$\{S, Se, Te\}, with 2H-t symmetry, the lighter chalcogen placed on the inner side of the nanotube in each case. These materials are chosen in the present work since they correspond to the full set of Janus TMD  nanotubes predicted to be stable from Kohn-Sham DFT calculations \cite{bolle2021structural}.

\begin{figure}[htbp!]
        \centering
        \includegraphics[width=0.9\textwidth]{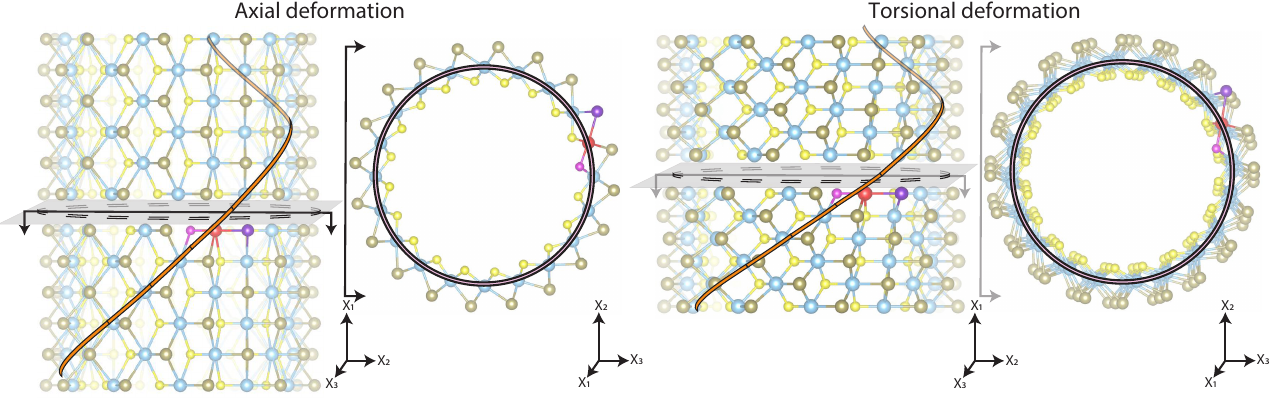}
        \caption{Illustration portraying the cyclic+helical symmetry inherent to  Janus TMD nanotubes subject to axial/torsional deformations, using a (10,10) 1T-o symmetry nanotube as the representative example  (structural model generated using VESTA \cite{momma2008vesta}). In particular,  the nanotube can be described by the symmetry operators and the positions of three atoms, e.g., metal and chalcogens colored red and violet/pink, respectively.  This structural and resultant electronic symmetry is exploited while performing ab initio calculations using the Cyclix-DFT code \cite{sharma2021real, xu2021sparc}.}
      \label{fig:illustration}
    \end{figure}

We perform simulations using the  massively parallel real-space DFT code SPARC \cite{xu2021sparc, ghosh2017sparc1, ghosh2017sparc2}. In particular, we employ the recently developed Cyclix-DFT feature \cite{sharma2021real}, which provides a cyclic+helical symmetry-adapted formulation \cite{sharma2021real, ghosh2019symmetry, banerjee2016cyclic} and implementation \cite{sharma2021real} of the Kohn-Sham problem,  enabling the simulation of Janus TMD nanotubes subject to axial and/or torsional deformations with only three atoms in the fundamental domain, i.e., one of each chemical element  (Fig.~\ref{fig:illustration}). This provides  tremendous reduction in the cost, making many of the calculations performed here tractable, e.g., a (50,50) TiSTe nanotube with diameter $\sim 10$ nm subject to a twist of $5\times10\textsuperscript{-4}$ rad/Bohr has $238,950$ atoms in the periodic unit cell, a system size that is clearly beyond the reach of traditional DFT formulations/implementations due to the cubic scaling diagonalization bottleneck. Notably, Cyclix-DFT is now a mature feature that has been validated and used for a number of  physical applications \cite{codony2021transversal, kumar2021flexoelectricity, sharma2021real, kumar2020bending, bhardwaj2021torsional, bhardwaj2021strain, bhardwaj2021elastic}.

We employ the Perdew–Burke–Ernzerhof (PBE) \cite{perdew1996generalized} exchange-correlation functional along with the optimized norm-conserving Vanderbilt (ONCV) pseudopotentials \cite{hamann2013optimized} from the SPMS \cite{spms} collection. The accuracy of these pseudopotentials for the study of Janus TMD nanotubes has been verified in recent work \cite{bhardwaj2021elastic}. Though PBE generally underpredicts the bandgap and is considered less reliable than hybrid functionals for computation of the band structure, the situation appears to be reversed for Janus TMD monolayers, a trend that is expected to hold for their nanotube counterparts as well. In particular, the bandgap for MoSSe monolayer computed here using PBE is in much better agreement with experiment than the Heyd–Scuseria–Ernzerhof (HSE) \cite{heyd2003hybrid} hybrid functional.
Specifically, the PBE and HSE values for MoSSe are 1.63 eV (Ref.~\cite{shi2018mechanical}: 1.56 eV) and 1.95/2.23 eV \cite{haastrup2018computational, evarestov2020first} respectively, with the experimental value being 1.48 eV \cite{zhang2017janus}. Note that for Janus TMD monolayers, only minor modifications to the band structure have been observed upon the inclusion of spin-orbit coupling \cite{haastrup2018computational}, hence we neglect it, expecting a similarly negligible effect for the nanotubes.

We use the above first principles framework to compute the change in bandgap and effective mass of charge carriers, i.e., electrons and holes, with axial and torsional deformations, for the aforementioned twenty-seven Janus TMD nanotubes, considering both the armchair and zigzag variants. The values of axial strain --- ratio of the change in length of nanotube to its original length --- and shear strain --- product of nanotube radius and applied twist per unit length --- are chosen to be in accordance with those applied in experiments \cite{kaplan2007mechanical, kaplan2006mechanical, levi2015nanotube, divon2017torsional, nagapriya2008torsional}. Additional details regarding the computation of the bandgap and  effective mass of charge carriers within the symmetry-adapted framework can be found in Ref.~\cite{sharma2021real}. All the numerical parameters for Cyclix-DFT, including the grid spacing for real-space discretization and Brillouin zone integration, radial vacuum, and structural relaxation tolerances are chosen such that the ground state Kohn-Sham energy is converged to within $10^{-4}$ Ha/atom, which results in the bandgap and charge carriers' effective mass being accurate to within 0.01 eV and 0.01 a.u., respectively.  
    
\section{Results and discussion} \label{Sec:Results}

We now present and discuss the electronic response of the aforementioned Janus TMD nanotubes to axial/torsional deformations, from first principles,  using the symmetry-adapted Kohn-Sham DFT framework described in the previous section. The raw data for the bandgaps and effective masses is  available in the Supplementary Material. Wherever available, we compare the results obtained here with those in literature. 

\begin{figure}[h!]
        \centering
        \includegraphics[width=0.9\textwidth]{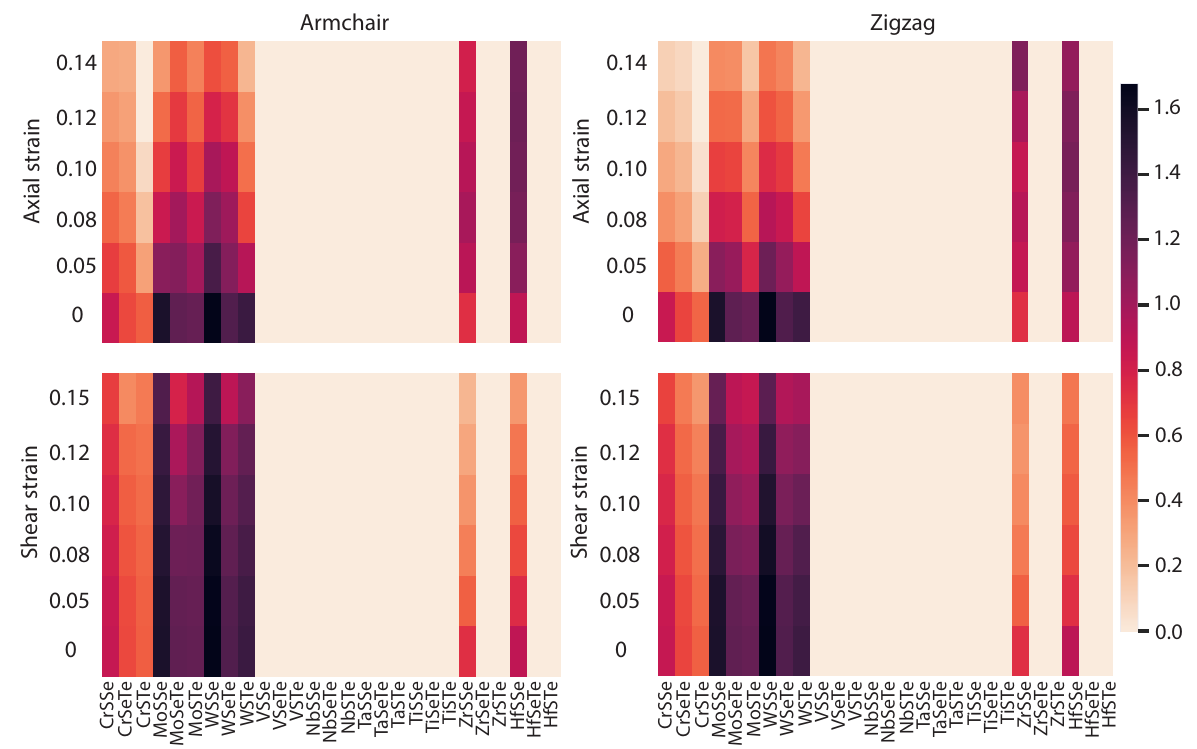}
        \caption{Variation of bandgap  with axial and torsional deformations for the twenty-seven select armchair and zigzag Janus TMD nanotubes.}
      \label{fig:heatbandgap}
    \end{figure}

First, we summarize the variation of bandgap with axial and shear strains in Fig.~\ref{fig:heatbandgap}. We observe that the undeformed CrSSe, CrSeTe, CrSTe, MoSSe, MoSeTe, MoSTe, WSSe, WSeTe, WSTe, ZrSSe, and HfSSe nanotubes are semiconducting, while the rest are metallic. In particular, bandgaps for MoSSe and MoSTe are in agreement with previous DFT studies \cite{wu2018tuning, tang2018janus, mikkelsen2021band, zhao2015ultra}. Upon deformation, the metallic nanotubes continue to remain metallic (behavior that distinguishes them from carbon nanotubes \cite{minot2003tuning, sharma2021real}), whereas semiconducting nanotubes undergo a monotonic decrease in bandgap, the exceptions being HfSSe: bandgap increases with axial strain, and ZrSSe: bandgap first increases and then decreases with axial strain. Such tunability of bandgaps has a number of applications in nanotechnology, e.g., mechanical sensors \cite{li2016low, sorkin2014nanoscale, oshima2020geometrical}. Notably, both armchair and zigzag CrSTe nanotubes go through a semiconductor to metal transition within the range of axial strains considered. Indeed, for large enough axial/torsional deformations, such semiconductor to metal transitions are likely to occur for all  the semiconducting nanotubes, though stability considerations become particularly important and therefore need to be considered/addressed.
 
\begin{figure}[h!]
        \centering
        \includegraphics[width=0.9\textwidth]{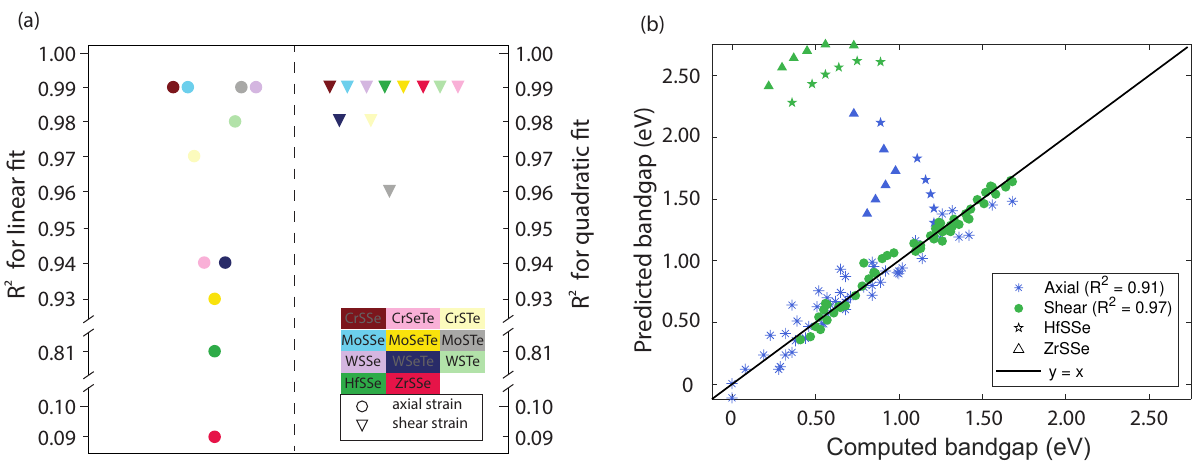}
        \caption{(a) The coefficient of determination of the linear regression (R\textsuperscript{2}) for the linear and quadratic fits of the bandgap vs. axial and shear strain, respectively. (b) The computed bandgap vs. that  predicted using the linear regression model. The two outliers: HfSSe and ZrSSe, are not included in the regression. The systems under consideration are the eleven semiconducting armchair Janus TMD nanotubes, with results for the zigzag variants being nearly identical to those presented here.}
      \label{fig:regression}
    \end{figure}

Observing a linear and quadratic variation of the semiconducting Janus TMD nanotubes' bandgap with axial and shear strains, respectively, we present results of the corresponding fits in Fig.~\ref{fig:regression}. It is clear that the fits are  excellent, with the exception  being ZrSSe, where there is a non-monotonic variation of the bandgap, as discussed above.  The linear dependence of the bandgap on axial strain, which has also been observed for the following TMD nanotubes: MoS\textsubscript{2}, MoSe\textsubscript{2}, WS\textsubscript{2},  WSe\textsubscript{2}, and CrS\textsubscript{2} \cite{zibouche2014electromechanical, li2014strain, ghorbani2013electromechanics, wang2016strain}, is consistent with previous results for  MoSSe \cite{luo2019electronic, wang2018mechanical}. In particular,  the slope computed here is $-8.7$, which is in good agreement with the slope $-7.1$ predicted by Ref.~\cite{wang2018mechanical}. Note that the exact values cannot be compared, since Refs.~\cite{luo2019electronic, wang2018mechanical} choose different diameters than the equilibrium values uesd here. To develop a simple model for the bandgap variation, we also perform linear regression with the features being metal-chalcogen bond lengths, chalcogens' electronegativity difference, axial strain, shear strain, and square of the shear strain, the results of which are presented in Fig.~\ref{fig:regression}. It is clear that, again with the exception of HfSSe and ZrSSe,  there is very good agreement between the computed values and those predicted by the regression model, suggesting the importance of the metal-chalcogen bond lengths and the chalcogens' electronegativity difference in determining the bandgap at any given deformation.

\begin{figure}[h!]
        \centering
        \includegraphics[width=0.9\textwidth]{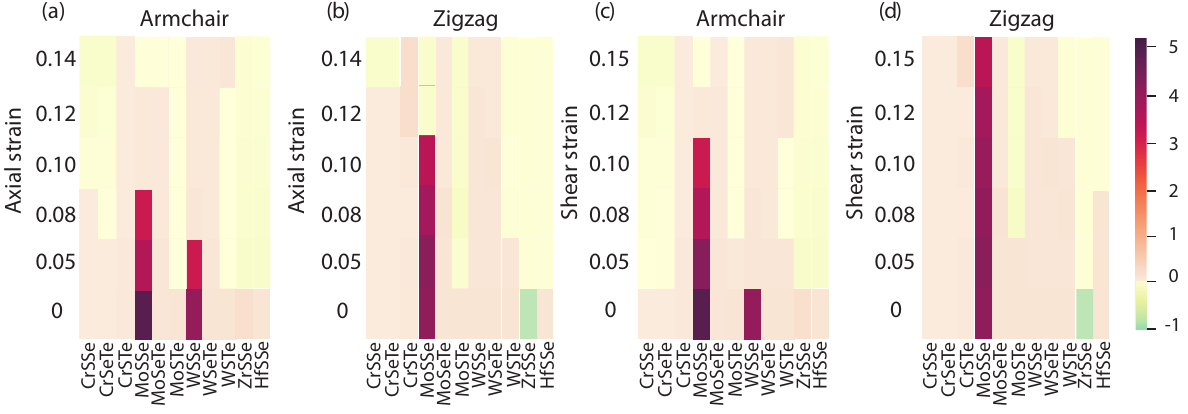}
        \caption{Variation of the effective mass of holes minus that of the electrons with axial and torsional deformations   for the eleven semiconducting Janus TMD nanotubes.}
      \label{fig:heatholemass}
    \end{figure}

Next, considering the eleven Janus TMD nanotubes (both armchair and zigzag configurations) that have determined to be semiconducting, we summarize the variation of the difference between effective mass of holes and electrons with deformations  in Fig.~\ref{fig:heatholemass}. Note that this difference in the effective masses can be used to classify whether the material is a n-type or p-type semiconductor, i.e., if the effective mass of hole is greater than of the electron, then holes have lower mobility, resulting in n-type semiconductor behavior, and vice versa \cite{hautier2013identification}. Here, all the nanotubes are n-type semiconductors in the undeformed configuration, the exception again being ZrSSe. The application of deformation is accompanied by the continuous decrease and increase in the effective mass of holes and electrons, respectively, culminating in an n-type to p-type semiconductor transition for armchair CrSSe, CrSeTe, MoSSe, MoSeTe, MoSTe, WSTe, ZrSSe, and HfSSe, and zigzag CrSSe, CrSeTe, MoSSe, MoSTe, WSTe, and HfSSe nanotubes with axial deformation; and armchair CrSSe, CrSeTe, MoSSe, MoSTe, WSTe, ZrSSe, and HfSSe, and zigzag MoSTe, WSTe, and HfSSe nanotubes with torsional deformation, for the range of strains considered.  Indeed, similar to the case of bandgaps, larger strains are expected to introduce transitions for the remaining  nanotubes as well, however stability considerations are likely to become important. Such transitions have applications as semiconductor switches \cite{nilges2009reversible, zhang2016switching, hiramatsu2007heavy, chen2015p, wen2018pressure, naab2013high}.

\begin{figure}[h!]
        \centering
        \includegraphics[width=0.9\textwidth]{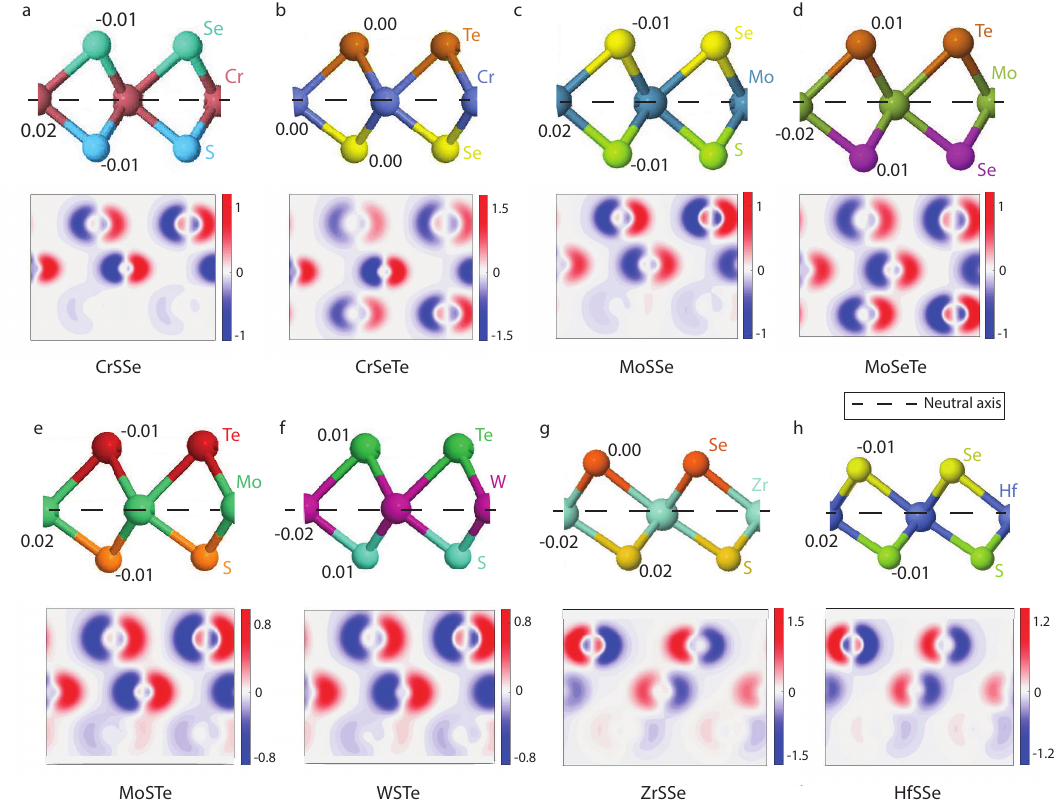}
        \caption{Contours of difference in electron density (integrated along the $x_1$ direction) between the  strained --- smallest value at which n-type to p-type semiconductor transition occurs --- and unstrained armchair nanotube configurations. The top four panels (a, b, c, and d) correspond to axial deformations, and the bottom four panels (e, f, g, and h) correspond to torsional deformations. Contour plots are on $x_2 x_3$-plane w.r.t their monolayer flat sheet configuration. The numbers listed next to the atoms correspond to the amount of charge transfer, as determined by Bader analysis \cite{bader1981quantum, tang2009grid}.}
      \label{fig:contour}
    \end{figure}

To gain further physical understanding into the results presented above, considering the semiconducting Janus TMD nanotubes that undergo n-type to p-type transitions, we choose eight representative ones, i.e.,  four each for transitions resulting from axial and torsional deformations, and plot their electron density difference contours between the strained (smallest value at which transition has occurred) and unstrained nanotube configurations in Fig.~\ref{fig:contour}.   In addition, we utilize Bader analysis \cite{bader1981quantum, tang2009grid} to determine the amount of metal-chalcogen charge transfer that has occurred between the undeformed and deformed nanotube configurations, results of which are also presented in Fig.~\ref{fig:contour}. We find that there is a minimal change in the Bader charge of atoms upon deformation, suggesting that the nature of the bonding between the atoms remains unaltered. However, there is a significant change in the electron density contours, leading to the conclusion that the observed behavior is a consequence of the rehybridization of orbitals.

\section{Concluding remarks} \label{Sec:Conclusions}
We have investigated the electronic response of Janus TMD nanotubes to mechanical deformations using ab-intio simulations. In particular, considering the full set of twenty-seven such nanotubes predicted to be stable in literature, we have performed Kohn-Sham calculations to determine the change in bandgap and charge carriers' effective mass upon axial and torsional deformations, for both the armchair and zigzag variants. We have found that the metallic nanotubes continue to remain metallic even upon deformation, whereas  semiconducting nanotubes' bandgap generally decreases with axial and shear strains, in linear and quadratic fashion, respectively, leading to  semiconductor--metal transitions. In addition, we have observed a continual decrease and increase in mass of holes and electrons, respectively, with strains, leading to transitions from n-type to p-type semiconducting behavior.  We have used electronic and charge transfer analysis to show that the response is determined by the rehybridization of orbitals, instead of charge transfer between the atoms. Overall, mechanical deformations represent powerful tools for tailoring the electronic response of semiconducting Janus TMD nanotubes, which has a number of applications, including devices such as semiconductor switches and mechanical sensors.

The stability of Janus TMD nanotubes, as determined from phonon analyses, in both undeformed and deformed configurations, is a worthy topic for future research. In addition, the thermal and optical response of these nanotubes to mechanical deformations presents itself as another worthy subject for future investigation. Finally, the flexoelectric properties of Janus TMD nanotubes is an interesting problem that is also worthy of pursuit.

\section*{Acknowledgements} 
The authors gratefully acknowledge the support of the US National Science Foundation (CAREER-1553212).


\vspace{-1mm}

\begin{thebibliography}{10}

\bibitem{iijima1991helical}
S~Iijima.
\newblock {Helical microtubules of graphitic carbon}.
\newblock {\em Nature}, 354(6348):56--58, 1991.

\bibitem{tenne2003advances}
R~Tenne.
\newblock {Advances in the synthesis of inorganic nanotubes and fullerene-like
  nanoparticles}.
\newblock {\em Angewandte Chemie International Edition}, 42(42):5124--5132,
  2003.

\bibitem{rao2003inorganic}
C~N~R Rao and M~Nath.
\newblock {Inorganic nanotubes}.
\newblock In {\em Advances In Chemistry: A Selection of CNR Rao's Publications
  (1994--2003)}, pages 310--333. World Scientific, 2003.

\bibitem{serra2019overview}
M~Serra, R~Arenal, and R~Tenne.
\newblock {An overview of the recent advances in inorganic nanotubes}.
\newblock {\em Nanoscale}, 11(17):8073--8090, 2019.

\bibitem{haastrup2018computational}
S~Haastrup, M~Strange, M~Pandey, T~Deilmann, P~S Schmidt, N~F Hinsche, M~N
  Gjerding, D~Torelli, P~M Larsen, A~C Riis-Jensen, et~al.
\newblock {The Computational 2D Materials Database: high-throughput modeling
  and discovery of atomically thin crystals}.
\newblock {\em 2D Materials}, 5(4):042002, 2018.

\bibitem{zhou20192dmatpedia}
J~Zhou, L~Shen, M~D Costa, K~A Persson, Shyue~P Ong, P~Huck, Y~Lu, X~Ma,
  Y~Chen, H~Tang, et~al.
\newblock {2DMatPedia, an open computational database of two-dimensional
  materials from top-down and bottom-up approaches}.
\newblock {\em Scientific Data}, 6(1):1--10, 2019.

\bibitem{gjerding2021recent}
M~N Gjerding, A~Taghizadeh, A~Rasmussen, S~Ali, F~Bertoldo, T~Deilmann, U~P
  Holguin, N~R Kn{\o}sgaard, M~Kruse, S~Manti, et~al.
\newblock {Recent Progress of the Computational 2D Materials Database (C2DB)}.
\newblock {\em 2D Materials}, 8(4):044002, 2021.

\bibitem{kumar2020bending}
S~Kumar and P~Suryanarayana.
\newblock {Bending moduli for forty-four select atomic monolayers from first
  principles}.
\newblock {\em Nanotechnology}, 31(43):43LT01, 2020.

\bibitem{yagmurcukardes2020quantum}
M~Yagmurcukardes, Y~Qin, S~Ozen, M~Sayyad, Fran{\c{c}}ois~M Peeters, S~Tongay,
  and H~Sahin.
\newblock {Quantum properties and applications of 2D Janus crystals and their
  superlattices}.
\newblock {\em Applied Physics Reviews}, 7(1):011311, 2020.

\bibitem{xiong2018spontaneous}
Q-L Xiong, J~Zhou, J~Zhang, T~Kitamura, and Z-H Li.
\newblock {Spontaneous curling of freestanding Janus monolayer transition-metal
  dichalcogenides}.
\newblock {\em Physical Chemistry Chemical Physics}, 20(32):20988--20995, 2018.

\bibitem{bhardwaj2021elastic}
A~Bhardwaj and P~Suryanarayana.
\newblock {Elastic properties of Janus transition metal dichalcogenide
  nanotubes from first principles}.
\newblock {\em The European Physical Journal B}, 95(13):1--8, 2022.

\bibitem{lu2017janus}
A-Y Lu, H~Zhu, J~Xiao, C-P Chuu, Y~Han, M-H Chiu, C-C Cheng, C-W Yang, K-H Wei,
  Y~Yang, et~al.
\newblock {Janus monolayers of transition metal dichalcogenides}.
\newblock {\em Nature Nanotechnology}, 12(8):744--749, 2017.

\bibitem{zhang2017janus}
J~Zhang, S~Jia, I~Kholmanov, L~Dong, D~Er, W~Chen, H~Guo, Z~Jin, V~B Shenoy,
  L~Shi, et~al.
\newblock {Janus monolayer transition-metal dichalcogenides}.
\newblock {\em ACS Nano}, 11(8):8192--8198, 2017.

\bibitem{trivedi2020room}
D~B Trivedi, G~Turgut, Y~Qin, M~Y Sayyad, D~Hajra, M~Howell, L~Liu, S~Yang, N~H
  Patoary, H~Li, et~al.
\newblock {Room-Temperature Synthesis of 2D Janus Crystals and their
  Heterostructures}.
\newblock {\em Advanced Materials}, 32(50):2006320, 2020.

\bibitem{lin2020low}
Y-C Lin, C~Liu, Y~Yu, E~Zarkadoula, M~Yoon, A~A Puretzky, L~Liang, Y~Kong,
  Xand~Gu, A~Strasser, et~al.
\newblock {Low energy implantation into transition-metal dichalcogenide
  monolayers to form Janus structures}.
\newblock {\em ACS Nano}, 14(4):3896--3906, 2020.

\bibitem{tang2018janus}
Z-K Tang, B~Wen, M~Chen, and L-M Liu.
\newblock {Janus MoSSe nanotubes: tunable band gap and excellent optical
  properties for surface photocatalysis}.
\newblock {\em Advanced Theory and Simulations}, 1(10):1800082, 2018.

\bibitem{oshima2020geometrical}
S~Oshima, M~Toyoda, and S~Saito.
\newblock {Geometrical and electronic properties of unstrained and strained
  transition metal dichalcogenide nanotubes}.
\newblock {\em Physical Review Materials}, 4(2):026004, 2020.

\bibitem{xie2021theoretical}
S~Xie, H~Jin, Y~Wei, and S~Wei.
\newblock {Theoretical investigation on stability and electronic properties of
  Janus MoSSe nanotubes for optoelectronic applications}.
\newblock {\em Optik}, 227:166105, 2021.

\bibitem{ju2021tuning}
L~Ju, P~Liu, Y~Yang, L~Shi, G~Yang, and L~Sun.
\newblock {Tuning the photocatalytic water-splitting performance with the
  adjustment of diameter in an armchair WSSe nanotube}.
\newblock {\em Journal of Energy Chemistry}, 61:228--235, 2021.

\bibitem{ju2021rolling}
L~Ju, J~Qin, G~Shi, Land~Yang, J~Zhang, and L~Sun.
\newblock {Rolling the WSSe bilayer into double-walled nanotube for the
  enhanced photocatalytic water-splitting performance}.
\newblock {\em Nanomaterials}, 11(3):705, 2021.

\bibitem{zhang2019mosse}
S~Zhang, H~Jin, C~Long, T~Wang, R~Peng, B~Huang, and Y~Dai.
\newblock {MoSSe nanotube: a promising photocatalyst with an extremely long
  carrier lifetime}.
\newblock {\em Journal of Materials Chemistry A}, 7(13):7885--7890, 2019.

\bibitem{yudilevichself}
D~Yudilevich, R~Levi, I~Nevo, R~Tenne, A~Ya’akobovitz, and E~Joselevich.
\newblock {Self-sensing torsional resonators based on inorganic nanotubes}.
\newblock {\em ICME}, pages 1--4, 2018.

\bibitem{levi2015nanotube}
R~Levi, J~Garel, D~Teich, G~Seifert, R~Tenne, and E~Joselevich.
\newblock {Nanotube electromechanics beyond carbon: the case of
  WS\textsubscript{2}}.
\newblock {\em ACS Nano}, 9(12):12224--12232, 2015.

\bibitem{divon2017torsional}
Y~Divon, R~Levi, J~Garel, D~Golberg, R~Tenne, A~Ya’akobovitz, and
  E~Joselevich.
\newblock {Torsional resonators based on inorganic nanotubes}.
\newblock {\em Nano Letters}, 17(1):28--35, 2017.

\bibitem{barua2017nanostructured}
S~Barua, H~S Dutta, R~Gogoi, Sand~Devi, and R~Khan.
\newblock {Nanostructured MoS\textsubscript{2}-based advanced biosensors: a
  review}.
\newblock {\em ACS Applied Nano Materials}, 1(1):2--25, 2017.

\bibitem{li2016low}
B~L Li, J~Wang, H~L Zou, S~Garaj, C~T Lim, J~Xie, N~B Li, and D~T Leong.
\newblock {Low-dimensional transition metal dichalcogenide nanostructures based
  sensors}.
\newblock {\em Advanced Functional Materials}, 26(39):7034--7056, 2016.

\bibitem{sorkin2014nanoscale}
V~Sorkin, H~Pan, H~Shi, S~Y Quek, and Y~W Zhang.
\newblock {Nanoscale transition metal dichalcogenides: structures, properties,
  and applications}.
\newblock {\em Critical Reviews in Solid State and Materials Sciences},
  39(5):319--367, 2014.

\bibitem{nath2003superconducting}
M~Nath, S~Kar, A~K Raychaudhuri, and C~N~R Rao.
\newblock {Superconducting NbSe\textsubscript{2} nanostructures}.
\newblock {\em Chemical Physics Letters}, 368(5-6):690--695, 2003.

\bibitem{tsuneta2003formation}
T~Tsuneta, T~Toshima, K~Inagaki, T~Shibayama, S~Tanda, S~Uji, M~Ahlskog,
  P~Hakonen, and M~Paalanen.
\newblock {Formation of metallic NbSe\textsubscript{2} nanotubes and
  nanofibers}.
\newblock {\em Current Applied Physics}, 3(6):473--476, 2003.

\bibitem{wu2018tuning}
H-H Wu, Q~Meng, H~Huang, C~T Liu, and X-L Wang.
\newblock {Tuning the indirect--direct band gap transition in the
  MoS\textsubscript{2-x}Se\textsubscript{x} armchair nanotube by diameter
  modulation}.
\newblock {\em Physical Chemistry Chemical Physics}, 20(5):3608--3613, 2018.

\bibitem{mikkelsen2021band}
AE~G Mikkelsen, F~T B{\"o}lle, K~S Thygesen, T~Vegge, and I~E Castelli.
\newblock {Band structure of MoSTe Janus nanotubes}.
\newblock {\em Physical Review Materials}, 5(1):014002, 2021.

\bibitem{zhao2015ultra}
W~Zhao, Y~Li, W~Duan, and F~Ding.
\newblock {Ultra-stable small diameter hybrid transition metal dichalcogenide
  nanotubes X--M--Y (X, Y= S, Se, Te; M= Mo, W, Nb, Ta): a computational
  study}.
\newblock {\em Nanoscale}, 7(32):13586--13590, 2015.

\bibitem{tao2018band}
L~Tao, Y-Y Zhang, J~Sun, S~Du, and H-J Gao.
\newblock {Band engineering of double-wall Mo-based hybrid nanotubes}.
\newblock {\em Chinese Physics B}, 27(7):076104, 2018.

\bibitem{evarestov2020first}
R~A Evarestov, A~V Kovalenko, and A~V Bandura.
\newblock {First-principles study on stability, structural and electronic
  properties of monolayers and nanotubes based on pure
  Mo(W)S(Se)\textsubscript{2} and mixed (Janus)Mo(W)SSe dichalcogenides}.
\newblock {\em Physica E: Low-dimensional Systems and Nanostructures},
  115:113681, 2020.

\bibitem{bolle2021structural}
Felix~T B{\"o}lle, A~E~G Mikkelsen, K~S Thygesen, T~Vegge, and I~E Castelli.
\newblock {Structural and chemical mechanisms governing stability of inorganic
  Janus nanotubes}.
\newblock {\em npj Computational Materials}, 7(1):1--8, 2021.

\bibitem{wang2018mechanical}
Y~Z Wang, R~Huang, B~L Gao, G~Hu, F~Liang, and Y~L Ma.
\newblock {Mechanical and strain-tunable electronic properties of Janus MoSSe
  nanotubes}.
\newblock {\em Chalcogenide Letters}, 15(11):535--543, 2018.

\bibitem{luo2019electronic}
Y~F Luo, Y~Pang, M~Tang, Q~Song, and M~Wang.
\newblock {Electronic properties of Janus MoSSe nanotubes}.
\newblock {\em Computational Materials Science}, 156:315--320, 2019.

\bibitem{hohenberg1964inhomogeneous}
P~Hohenberg and W~Kohn.
\newblock {Inhomogeneous electron gas}.
\newblock {\em Physical Review}, 136(3B):B864, 1964.

\bibitem{kohn1965self}
W~Kohn and L~J Sham.
\newblock {Self-consistent equations including exchange and correlation
  effects}.
\newblock {\em Physical Review}, 140(4A):A1133, 1965.

\bibitem{momma2008vesta}
K~Momma and F~Izumi.
\newblock {VESTA: a three-dimensional visualization system for electronic and
  structural analysis}.
\newblock {\em Journal of Applied Crystallography}, 41(3):653--658, 2008.

\bibitem{sharma2021real}
A~Sharma and P~Suryanarayana.
\newblock {Real-space density functional theory adapted to cyclic and helical
  symmetry: Application to torsional deformation of carbon nanotubes}.
\newblock {\em Physical Review B}, 103(3):035101, 2021.

\bibitem{xu2021sparc}
Q~Xu, A~Sharma, B~Comer, H~Huang, E~Chow, A~J Medford, J~E Pask, and
  P~Suryanarayana.
\newblock SPARC: Simulation package for ab-initio real-space calculations.
\newblock {\em SoftwareX}, 15:100709, 2021.

\bibitem{ghosh2017sparc1}
S~Ghosh and P~Suryanarayana.
\newblock {SPARC: Accurate and efficient finite-difference formulation and
  parallel implementation of density functional theory: Isolated clusters}.
\newblock {\em Computer Physics Communications}, 212:189--204, 2017.

\bibitem{ghosh2017sparc2}
S~Ghosh and P~Suryanarayana.
\newblock {SPARC: Accurate and efficient finite-difference formulation and
  parallel implementation of Density Functional Theory: Extended systems}.
\newblock {\em Computer Physics Communications}, 216:109--125, 2017.

\bibitem{ghosh2019symmetry}
S~Ghosh, A~S Banerjee, and P~Suryanarayana.
\newblock {Symmetry-adapted real-space density functional theory for
  cylindrical geometries: Application to large group-IV nanotubes}.
\newblock {\em Physical Review B}, 100(12):125143, 2019.

\bibitem{banerjee2016cyclic}
A~S Banerjee and P~Suryanarayana.
\newblock {Cyclic density functional theory: A route to the first principles
  simulation of bending in nanostructures}.
\newblock {\em Journal of the Mechanics and Physics of Solids}, 96:605--631,
  2016.

\bibitem{codony2021transversal}
David Codony, Irene Arias, and Phanish Suryanarayana.
\newblock Transversal flexoelectric coefficient for nanostructures at finite
  deformations from first principles.
\newblock {\em Physical Review Materials}, 5(3):L030801, 2021.

\bibitem{kumar2021flexoelectricity}
S~Kumar, D~Codony, I~Arias, and P~Suryanarayana.
\newblock {Flexoelectricity in atomic monolayers from first principles}.
\newblock {\em Nanoscale}, 13(3):1600--1607, 2021.

\bibitem{bhardwaj2021torsional}
A~Bhardwaj, A~Sharma, and P~Suryanarayana.
\newblock Torsional moduli of transition metal dichalcogenide nanotubes from
  first principles.
\newblock {\em Nanotechnology}, 32(28):28LT02, 2021.

\bibitem{bhardwaj2021strain}
A~Bhardwaj, A~Sharma, and P~Suryanarayana.
\newblock {Torsional strain engineering of transition metal dichalcogenide
  nanotubes: An ab initio study}.
\newblock {\em Nanotechnology}, 32(47):47LT01, 2021.

\bibitem{perdew1996generalized}
J~P Perdew, K~Burke, and M~Ernzerhof.
\newblock {Generalized gradient approximation made simple}.
\newblock {\em Physical Review Letters}, 77(18):3865, 1996.

\bibitem{hamann2013optimized}
D~R Hamann.
\newblock {Optimized norm-conserving Vanderbilt pseudopotentials}.
\newblock {\em Physical Review B}, 88(8):085117, 2013.

\bibitem{spms}
M.~F. Shojaei, J.~E. Pask, A.~J. Medford, and P.~Suryanarayana.
\newblock {Soft and transferable pseudopotentials from multi-objective optimization}.
\newblock {\em arXiv preprint arXiv:2209.09806}, 2022.

\bibitem{heyd2003hybrid}
J~Heyd, G~E Scuseria, and M~Ernzerhof.
\newblock {Hybrid functionals based on a screened Coulomb potential}.
\newblock {\em The Journal of Chemical Physics}, 118(18):8207--8215, 2003.

\bibitem{shi2018mechanical}
W~Shi and Z~Wang.
\newblock {Mechanical and electronic properties of Janus monolayer transition
  metal dichalcogenides}.
\newblock {\em Journal of Physics: Condensed Matter}, 30(21):215301, 2018.

\bibitem{kaplan2007mechanical}
I~Kaplan-Ashiri and R~Tenne.
\newblock {Mechanical properties of WS\textsubscript{2} nanotubes}.
\newblock {\em Journal of Cluster Science}, 18(3):549--563, 2007.

\bibitem{kaplan2006mechanical}
I~Kaplan-Ashiri, S~R Cohen, K~Gartsman, V~Ivanovskaya, T~Heine, G~Seifert,
  I~Wiesel, H~D Wagner, and R~Tenne.
\newblock {On the mechanical behavior of WS\textsubscript{2} nanotubes under
  axial tension and compression}.
\newblock {\em Proceedings of the National Academy of Sciences},
  103(3):523--528, 2006.

\bibitem{nagapriya2008torsional}
K~S Nagapriya, O~Goldbart, I~Kaplan-Ashiri, G~Seifert, R~Tenne, and
  E~Joselevich.
\newblock {Torsional stick-slip behavior in WS\textsubscript{2} nanotubes}.
\newblock {\em Physical Review Letters}, 101(19):195501, 2008.

\bibitem{minot2003tuning}
E~D Minot, Y~Yaish, V~Sazonova, J-Y Park, M~Brink, and P~L McEuen.
\newblock {Tuning carbon nanotube band gaps with strain}.
\newblock {\em Physical Review Letters}, 90(15):156401, 2003.

\bibitem{zibouche2014electromechanical}
N~Zibouche, M~Ghorbani-Asl, T~Heine, and A~Kuc.
\newblock {Electromechanical properties of small transition-metal
  dichalcogenide nanotubes}.
\newblock {\em Inorganics}, 2(2):155--167, 2014.

\bibitem{li2014strain}
W~Li, G~Zhang, M~Guo, and Y-W Zhang.
\newblock {Strain-tunable electronic and transport properties of
  MoS\textsubscript{2} nanotubes}.
\newblock {\em Nano Research}, 7(4):518--527, 2014.

\bibitem{ghorbani2013electromechanics}
M~Ghorbani-Asl, N~Zibouche, M~Wahiduzzaman, A~F Oliveira, A~Kuc, and T~Heine.
\newblock {Electromechanics in MoS\textsubscript{2} and WS\textsubscript{2}:
  nanotubes vs. monolayers}.
\newblock {\em Scientific Reports}, 3:2961, 2013.

\bibitem{wang2016strain}
Y~Z Wang, R~Huang, X~Q Wang, Q~F Zhang, B~L Gao, L~Zhou, and G~Hua.
\newblock {Strain-tunable electronic properties of CrS\textsubscript{2}
  nanotubes}.
\newblock {\em Chalcogenide Letters}, 13(7):301--307, 2016.

\bibitem{hautier2013identification}
G~Hautier, A~Miglio, G~Ceder, G-M Rignanese, and X~Gonze.
\newblock {Identification and design principles of low hole effective mass
  p-type transparent conducting oxides}.
\newblock {\em Nature Communications}, 4(1):1--7, 2013.

\bibitem{nilges2009reversible}
T~Nilges, S~Lange, M~Bawohl, J~M Deckwart, M~Janssen, H-D Wiemh{\"o}fer,
  R~Decourt, B~Chevalier, J~Vannahme, H~Eckert, et~al.
\newblock {Reversible switching between p-and n-type conduction in the
  semiconductor Ag\textsubscript{10}Te\textsubscript{4}Br\textsubscript{3}}.
\newblock {\em Nature Materials}, 8(2):101--108, 2009.

\bibitem{zhang2016switching}
J~Zhang, P~Gu, G~Long, R~Ganguly, Y~Li, N~Aratani, H~Yamada, and Q~Zhang.
\newblock {Switching charge-transfer characteristics from p-type to n-type
  through molecular “doping”(co-crystallization)}.
\newblock {\em Chemical Science}, 7(6):3851--3856, 2016.

\bibitem{hiramatsu2007heavy}
H~Hiramatsu, K~Ueda, H~Ohta, M~Hirano, M~Kikuchi, H~Yanagi, T~Kamiya, and
  H~Hosono.
\newblock {Heavy hole doping of epitaxial thin films of a wide gap p-type
  semiconductor, LaCuOSe, and analysis of the effective mass}.
\newblock {\em Applied Physics Letters}, 91(1):012104, 2007.

\bibitem{chen2015p}
L~Chen, J~Yang, S~Klaus, L~J Lee, R~Woods-Robinson, J~Ma, Y~Lum, J~K Cooper,
  F~M Toma, L-W Wang, et~al.
\newblock {p-Type transparent conducting oxide/n-type semiconductor
  heterojunctions for efficient and stable solar water oxidation}.
\newblock {\em Journal of the American Chemical Society}, 137(30):9595--9603,
  2015.

\bibitem{wen2018pressure}
T~Wen, Y~Wang, N~Li, Q~Zhang, Y~Zhao, W~Yang, Y~Zhao, and H-K Mao.
\newblock {Pressure-driven reversible switching between n-and p-Type conduction
  in chalcopyrite CuFeS\textsubscript{2}}.
\newblock {\em Journal of the American Chemical Society}, 141(1):505--510,
  2018.

\bibitem{naab2013high}
B~D Naab, S~Himmelberger, Y~Diao, K~Vandewal, P~Wei, B~Lussem, A~Salleo, and
  Z~Bao.
\newblock {High mobility n-type transistors based on solution-sheared doped
  6,13-Bis (triisopropylsilylethynyl) pentacene thin films}.
\newblock {\em Advanced Materials}, 25(33):4663--4667, 2013.

\bibitem{bader1981quantum}
R~F~W Bader and TT~Nguyen-Dang.
\newblock {Quantum theory of atoms in molecules--Dalton revisited}.
\newblock In {\em Advances in Quantum Chemistry}, volume~14, pages 63--124.
  Elsevier, 1981.

\bibitem{tang2009grid}
W~Tang, E~Sanville, and G~Henkelman.
\newblock A grid-based bader analysis algorithm without lattice bias.
\newblock {\em Journal of Physics: Condensed Matter}, 21(8):084204, 2009.

\end{thebibliography}
\end{document}